\documentclass[aps,prb,reprint,amsmath,amssymb,superscriptaddress,showpacs,floatfix]{revtex4-2}

\usepackage{mathtools}
\usepackage{bm}
\usepackage{mleftright}
\mleftright

\usepackage[dvipsnames,table,hyperref,rgb]{xcolor}
\definecolor{linkcolor}{rgb}{0.16, 0.32, 0.75}
\definecolor{citecolor}{rgb}{0.0, 0.5, 0.0}
\definecolor{urlcolor}{rgb}{0.06, 0.46, 1.0}
\usepackage[%
  colorlinks,%
  linkcolor={linkcolor},%
  citecolor={citecolor},%
  urlcolor={urlcolor}%
]{hyperref}

\begin{document}

\title{Electron transport in a weakly disordered Weyl semimetal}

\author{M.\,E.\,Ismagambetov}
\affiliation{Max Planck Institute for Solid State Research, 70569 Stuttgart, Germany}

\author{P.\,M.\,Ostrovsky}
\affiliation{Max Planck Institute for Solid State Research, 70569 Stuttgart, Germany}
\affiliation{TKM, Karlsruhe Institute of Technology, 76131 Karlsruhe, Germany}

\begin{abstract}
Weyl semimetal is a solid material with isolated touching points between conduction and valence bands in its Brillouin zone---Weyl points. Low energy excitations near these points exhibit a linear dispersion and act as relativistic massless particles. Weyl points are stable topological objects robust with respect to most perturbations. We study effects of weak disorder on the spectral and transport properties of Weyl semimetals in the limit of low energies. We use a model of Gaussian white-noise potential and apply dimensional regularization scheme near three dimensions to treat divergent terms in the perturbation theory. In the framework of self-consistent Born approximation, we find closed expressions for the average density of states and conductivity. Both quantities are analytic functions in the limit of zero energy. We also include interference terms beyond the  self-consistent Born approximation up to the third order in the disorder strength. These interference corrections are stronger than the mean-field result and non-analytic as functions of energy. Our main result is the dependence of conductivity (in units $e^2/h$) on the electron concentration $\sigma = \sigma_0 - 0.891\, n^{1/3} + 0.115\, (n^{2/3}/\sigma_0) \ln|n|$.
\end{abstract}

\maketitle

\section{\label{sec:intro} Introduction}

In recent years, the model of massless Weyl fermions has attracted a great attention in condensed matter physics since the discovery of Weyl semimetals \cite{Balents11, Wan11, WSMreview, Vishwanath18}. The Weyl semimetal is a solid-state crystal having isolated touching points (Weyl nodes or Weyl points) between conduction and valence bands with low-energy excitations that act as relativistic massless fermions. These fermions are described by the standard Weyl Hamiltonian:
\begin{equation}
 H = v\, \bm{\sigma}\cdot\mathbf{p}
\end{equation}
in the vicinity of a Weyl point. Here $v$ is the fixed velocity of massless excitations and $\bm{\sigma} = \{\sigma_x,\, \sigma_y,\, \sigma_z\}$ is a vector of three standard Pauli matrices.

Each Weyl point has a definite chirality and can be represented as a magnetic monopole, which is a source or sink of Berry flux in the momentum space. It is known that the net chirality of all touching points in the Brillouin zone must vanish \cite{Nielsen83} hence there is always an even number of Weyl nodes in the spectrum. As a consequence of the Nielsen-Ninomiya theorem \cite{Nielsen83}, the topological nature of such nodes protects them from opening a gap. Weak perturbations can only shift a Weyl point in momentum or energy while preserving its topological nature. The only possible way to open a spectral gap is by coupling two distant Weyl points. Hence the Weyl semimetal properties are robust with respect to small and smooth perturbations \cite{Solomon23}.

An important topological feature of any Weyl se\-mi\-me\-tal is the existence of low energy surface states that form a Fermi arc \cite{Wan11, Belopolski15} connecting projections of Weyl points on the crystal surface. These states exist on the background of excitationless bulk spectrum since the low energy excitations in the bulk occur only near isolated Weyl points. Fermi arc surface states are very well visible in the momentum-resolved spectroscopic measurements and are used as a hallmark for detecting Weyl semimetals. Recent experiments \cite{Lv15, Weng15} have proposed several candidate materials for Weyl semimetals: TaAs, TaP, NbAs, NbP. So far, the best candidate TaAs, studied in Ref.\ \cite{Weng15}, has shown 12 pairs of Weyl nodes.

One of the hottest debated topics in the theory of Weyl semimetals is the proposed quantum phase transition in the low-energy behavior of the density of states. Early numerical simulations of disordered Weyl semimetals \cite{Sbierski14, Trescher17, DasSarma16-2, Bera16, Roy18, Kobayashi14} have suggested that the density of states at zero energy undergoes a second-order transition from zero to a nonzero value when the strength of potential disorder exceeds a certain threshold. This conclusion is supported by some theoretical analysis. It was shown that a standard perturbation theory in weak disorder \cite{Syzranov15-1, Syzranov15-2, Roy16, Syzranov16, Louvet16, Erratum} developed near the dimension $d = 2$ and then continued to $d = 3$ indeed suggests vanishing density of states at zero energy. Alternative consideration in the framework of the nonlinear sigma model \cite{Frad86-1, Frad86-2, Altland15} provided a similar result. At the same time, it is quite clear that even an extremely small probability of a disorder realization that localizes an electron at zero energy is enough to disprove the proposed phase transition \cite{Nandishore14, Holder17, DasSarma16-1, Wilson20, Gurarie17}. Another nonperturbative approach \cite{Buchhold18} has suggested that for a broad class of ``optimal'' fluctuations of disorder potential the density of states is still exactly vanishing. Let us point out that behavior of DOS depends on the type of impurities, e.g. in the case of spherical impurities DOS at Weyl point should be nonzero  due to resonant scattering \cite{Pires21,Holder17}.Finally, diagrammatic calculations directly in 3D with Gaussian disorder and different types of momentum cutoff regularization \cite{Ominato14, Klier19} have shown the existence of the phase transition within the mean-field approximation. It can be thus claimed that although a true phase transition in the density of states is hardly possible, there is a very sharp crossover from algebraic to exponentially small density of states as a function of disorder strength \cite{Pix21}.

In the present work, we will study disorder effects in a Weyl semimetal in the limit of weak disorder only. Moreover, we will adopt the most standard Gaussian white-noise model of disorder similar to the one considered in Ref.\ \cite{Klier19}. Technically, this problem is identical to the well-studied Gross-Neveu model \cite{Gross74} in the Euclidean (imaginary time) representation with zero mass. Critical dimension of the Gross-Neveu model is $d = 2$. Close to this dimension, it is possible to develop a standard perturbative renormalization group approach to take into account logarithmically divergent diagrams. Such calculations were carried out up to the four loop order in Refs.\ \cite{Bondi90, Gracey16}. However, disordered Weyl semimetals correspond to the 3D version of the Gross-Neveu model where strong ultraviolet divergences make the renormalization group analysis problematic \cite{Rosenstein89}.

This work is devoted to 3D Weyl semimetals with weak Gaussian white-noise disorder. As was already pointed out, this model suffers from strong ultraviolet divergences and hence should be properly regularized. We will apply a standard technique of dimensional regularization \cite{ZJustin02} and allow the dimensionality of the system to deviate from $d = 3$. But contrary to the previous works \cite{Syzranov15-1, Syzranov15-2, Roy16, Syzranov16}, we will not perform any expansion near $d = 2$. Instead, we will consider all disorder corrections in a completely arbitrary dimension and do an exact analytic continuation of the results from the region $d < 2$ where ultraviolet divergences are absent to $d = 3$. Since the Gross-Neveu model is not renormalizable in $d = 3$, we will not develop any effective field theory description but instead consider disorder corrections directly to the observable quantities: density of states and conductivity. A calculation of weak disorder effects on the optical conductivity has been performed previously in Ref.\ \cite{DasSarma16-3}. In this work, we focus instead on the dc conductivity regime.

Our approach is quite similar to the standard problem of a conventional 3D metal with parabolic spectrum and Gaussian disorder. Detailed analysis of weak disorder effects in this model was performed in Refs.\ \cite{Kirk86, Wysokinski95}. It was shown that the model exhibits three types of corrections to the conductivity. First, there are corrections of relative strength $\propto (E \tau)^{-1}$, where $E$ is the Fermi energy and $\tau$ is the mean free time of the electrons. Second, there are weaker logarithmic corrections $\propto (E \tau)^{-2} \ln(E \tau)$. Such logarithms do not represent truly divergent contributions but require an accurate analysis of the diagrams on a ballistic scale. Finally, there is another logarithmic contribution $\propto (E \tau)^{-2} \ln(E/\Delta)$, where $\Delta$ is some ultraviolet energy cutoff scale related e.g.\ to the band width or to the lattice spacing. This type of a logarithmic divergence cannot be resolved in the low energy model with parabolic spectrum and requires an extra parameter $\Delta$. Divergent terms are present in the energy dependence of both the conductivity $\sigma$ and the total particle density $n$ but cancel out in their ratio, i.e. mobility of the metal $\mu = \sigma/(en)$.

The problem of a disordered Weyl semimetal considered in this work is technically more challenging. We will encounter stronger ultraviolet divergences that can be absorbed into redefinition of model parameters (Fermi energy, disorder strength etc). Such finite renormalizations are automatically taken into account by the dimensional regularization scheme \cite{ZJustin02}. The remaining logarithmic divergences will be cut at the scale $\Delta$. They do not cancel out in any observable quantity and will constitute an important part of our results. Let us emphasize that we develop a perturbative approach applicable in a relatively broad range of parameters (such as Fermi energy $E$, mean free time $\tau$ and UV cutoff $\Delta$). At the same time, possible nonperturbative effects, that are widely discussed in the context of Weyl semimetals \cite{Sbierski14, Trescher17, DasSarma16-2, Bera16, Roy18,Syzranov15-1, Syzranov15-2, Roy16, Syzranov16, Louvet16, Erratum, Altland15, Nandishore14, Holder17, DasSarma16-1, Buchhold18, Ominato14, Klier19, Pix21}, may occur only in an exponentially narrow parameter range near the Weyl point and will be disregarded in this work. We provide detailed estimates of the applicability of our approach in the Discussion section at the end of the paper.

The structure of the work is the following. In Sec.\ \ref{sec:Statement}, we formulate the problem and explain some details of the dimensional regularization scheme. Sec.\ \ref{sec:SCBA} contains a mean-field calculation of the density of states and conductivity based on the self-consistent Born approximation. This approach is similar to Ref.\ \cite{Klier19}. In Sec.\ \ref{sec:beyond}, we calculate interference contributions due to diagrams with two and three intersecting impurity lines. We show that these diagrams provide non-analytic corrections to the observable quantities. Main results are summarized and discussed in Sec.\ \ref{sec:conclusion}. Technically intricate details of the calculation of polarization operators in an arbitrary dimension are outlined in Appendix \ref{App}.

\section{\label{sec:Statement} Statement of the problem}

We consider a standard model of a single-node Weyl semimetal in the presence of potential disorder described by the following Hamiltonian:
\begin{equation}\label{ham}
 H = \bm{\sigma}\cdot \mathbf{p} + V(\mathbf{r}).
\end{equation}
For simplicity, we set the velocity of electrons to unity. In a real Weyl semimetal the number of Weyl nodes is at least two. Our model implies that disorder scattering between these nodes is negligible or, in other words, the disorder potential is smooth on the scale of inverse distance between Weyl nodes in momentum space. Linear momentum dependence of the Hamiltonian is of course also an approximation, it is valid only at low enough energies.

We assume that the random disorder potential obeys the standard Gaussian white-noise statistics
\begin{equation}\label{Gauss}
    \bigl< V(\mathbf{r}) \bigr> = 0,
    \qquad
    \bigl< V(\mathbf{r}) V(\mathbf{r'}) \bigr> = 2\pi^2 \alpha\: \delta(\mathbf{r - r'}).
\end{equation}
Disorder strength is characterized by a single parameter $\alpha$, that has a dimension of inverse energy for a 3D problem. The only dimensionless small parameter of our model is $\alpha E \ll 1$, where $E$ is the Fermi energy measured from the Weyl point. All observable quantities are even functions of $E$ so, for definiteness, we assume $E > 0$.

We will calculate average density of states and conductivity of a Weyl semimetal perturbatively in $\alpha E$ using diagrammatic expansion. The unperturbed Green function of the Weyl Hamiltonian is
\begin{equation}\label{Green}
 G^{R/A}(E, \mathbf{p})
  = \frac{E + \bm{\sigma}\cdot \mathbf{p}}{(E \pm i0)^2 - p^2}.
\end{equation}
In some parts of the calculation it will be more convenient to use Matsubara representation with imaginary energy $E = i\epsilon$. Retarded/advanced functions are then retrieved by analytic continuation from positive/negative $\epsilon$. For diagrams that involve both types of Green functions we have to keep two different Matsubara energies. Hence it will be convenient to change the sign of the Matsubara energy for advanced Green functions such that analytic continuation is always performed in the upper complex half-plane of energy.
\begin{equation}\label{Matsubara}
 G(i\epsilon_R, \mathbf{p})
  = -\frac{i\epsilon_R + \bm{\sigma}\cdot \mathbf{p}}{\epsilon_R^2 + p^2},
 \quad
 G(-i\epsilon_A, \mathbf{p})
  = \frac{i\epsilon_A - \bm{\sigma}\cdot \mathbf{p}}{\epsilon_A^2 + p^2}.
\end{equation}
Where $\epsilon_{R/A} \mapsto \mp i E + 0$. When only retarded Green functions are used (e.g.\ in the calculation of the density of states) we will omit the index and write simply $\epsilon$ instead of $\epsilon_R$ since it obeys the usual convention for the Matsubara energy.

Calculating diagrams for a system with linear dispersion in 3D leads to strong ultraviolet divergences. In fact, the theory is free of such problems only in the dimension $d < 2$. To overcome this difficulty, we will use dimensional regularization scheme \cite{Peskin95, ZJustin02}. This means calculating every diagram in an arbitrary dimension $d$ and then performing analytic continuation of the result in the parameter $d$ from the domain of convergence $d < 2$ to the point $d = 3$. Ultraviolet divergence, that occurs in most diagrams at $d = 2$, manifests itself as a pole in the corresponding expression as a function of $d$. Analytic continuation allows us to bypass the pole and get some finite result for $d > 2$. We will comment on the physical meaning of this mathematical trick later. Let us stress once again, that unlike numerous other works we do not imply any kind of expansion in the vicinity of the critical dimension $d = 2$ but rather allow for arbitrary values of the parameter $d$.

The matrix-valued vector $\bm{\sigma}$ is generalized to $d$ dimensions by imposing anticommutation relations on its elements:
\begin{equation}\label{Clifford}
 \sigma_a \sigma_b + \sigma_b \sigma_a = 2\delta_{ab}, \qquad \delta_{aa} = d, \qquad \operatorname{tr} 1 = 2.
\end{equation}
Strictly speaking, the convention $\operatorname{tr} 1 = 2$ does not hold for arbitrary $d$. For example in $d = 4$, a minimal representation of the Dirac $\gamma$ matrices has the size $4$. However, we will apply dimensional scheme at $d = 3$ where standard Pauli matrices have dimension $2$. Hence, for our purposes, the relation $\operatorname{tr} 1 = 2$ is valid.

For the sake of convenience, we also generalize the disorder correlation function (\ref{Gauss}) to arbitrary dimension as follows:
\begin{equation}\label{VV}
 \bigl< V(\mathbf{r}) V(\mathbf{r'}) \bigr> = \frac{(2\pi)^d}{S_{d-1}} \alpha\: \delta(\mathbf{r - r'}).
\end{equation}
Here $S_{d-1}$ is the volume of a $(d-1)$-dimensional unit sphere. The parameter $\alpha$ itself has a dimension that depends on $d$: $[\alpha] = E^{2-d}$.

Our main goal is to calculate average density of states $\rho$ and conductivity $\sigma$ in the limit of small energy or, equivalently, weak disorder. We will use standard Kubo expressions for these quantities. Density of states is given by a single average Green function at coincident points:
\begin{equation}\label{rhotrace}
 \rho(E)
  = -\frac{1}{\pi} \operatorname{Im} \int (d^dp) \operatorname{tr} \bigl< G^R(E, \mathbf{p}) \bigr>.
\end{equation}
For a clean system, it is easy to calculate the density of states just by the area of the Fermi surface. In 3D, this yields
\begin{equation}\label{rho0}
 \rho_0(E)
  = \frac{4\pi E^2}{(2\pi)^3}
  = \frac{E^2}{2\pi^2}.
\end{equation}
The same result follows from Eq.\ (\ref{rhotrace}) with the Green function from Eq.\ (\ref{Green}).

Kubo formula for conductivity (measured in units $e^2/h$) is
\begin{equation}\label{Kubo}
\sigma(E)
  = \operatorname{Tr} \Bigl< \sigma_x G_R (E) \sigma_x G_A(E) \Bigr>.
\end{equation}
Here `Tr' implies the trace in the space of Pauli matrices and also an integral in the momentum space. Matrices $\sigma_x$ here represent velocity operators of the Weyl Hamiltonian, $\partial H/ \partial\mathbf{p} = \bm{\sigma}$.

Let us stress that Eq.\ (\ref{Kubo}) is correct only in the framework of the dimensional regularization scheme. A more general Kubo formula for conductivity includes also the terms with two retarded or two advanced Green functions. However, these terms vanish due to gauge invariance as long as the corresponding momentum integral is convergent.

\section{\label{sec:SCBA} Self-consistent Born approximation}

A standard approach in the theory of weakly disordered metals is the self-consistent Born approximation (SCBA). It takes into account only the diagrams with non-intersecting impurity lines. Such diagrams are most important because all the Green functions can be taken close to the mass shell (Fermi surface) without violating conservation of momentum. Other diagrams, with intersection of impurity lines acquire an extra small factor in the limit $E \tau \gg 1$.

To estimate the disorder scattering rate in our model, we can apply the Fermi golden rule $1/\tau \sim \alpha \rho(E)$. With the density of states (\ref{rho0}), this yields
\begin{equation}\label{Etau}
 E \tau \sim \frac{1}{\alpha E} \gg 1.
\end{equation}
Hence the criterion of weak disorder $\alpha E\ll1$ also implies validity of SCBA. Let us stress quite a counterintuitive feature of the Weyl semimetal model: disorder effects get weaker with lowering the energy and shrinking of the Fermi surface. For most other common Hamiltonians the situation is opposite: parameter $E \tau$ grows with increasing energy. Weyl semimetals are special in this respect because their density of states has a relatively strong energy dependence with a soft gap at $E = 0$.

In the framework of SCBA, the electron Green function averaged over disorder realizations acquires a self energy which is independent of momentum
\begin{equation}\label{GSigma}
 G(i\epsilon, \mathbf{p})
  = \bigl[ i\epsilon - \bm{\sigma}\cdot \mathbf{p} - \Sigma(i\epsilon) \bigr]^{-1}.
\end{equation}
The self energy $\Sigma(i\epsilon)$ is identified with the simplest first-order Born diagram involving a single impurity line and a single Green function (\ref{GSigma}) that involves the same self energy. This yields the SCBA self-consistency equation for $\Sigma(i\epsilon)$ that automatically takes into account all self-energy diagrams with non-intersecting impurity lines. Explicitly, the equation is
\begin{equation}\label{SCBA}
 \Sigma
  = \frac{(2\pi)^d}{S_{d-1}}\: \alpha \int (d^d p)\; G(i\epsilon, \mathbf{p})
  = -\frac{i\pi\alpha (\epsilon+i\Sigma)^{d-1}}{2 \sin(\pi d/2)}.
\end{equation}
While the momentum integral here converges only for $d < 2$, we can perform analytic continuation of the result directly to the point $d = 3$. This way we arrive at a simple quadratic equation for the self energy that can be readily solved:
\begin{equation}\label{SCBA3D}
 \Sigma
  = \frac{i\pi\alpha}{2} (\epsilon + i \Sigma)^2
  = i\epsilon + \frac{i}{\pi\alpha} \Bigl( 1 - \sqrt{1 + 2\pi\alpha\epsilon} \Bigr).
\end{equation}
This equation and its solution can be directly compared to a similar SCBA equation studied in Ref.\ \cite{Klier19}. In that work, the divergence of the integral (\ref{SCBA}) was regulated by limiting the momenta $0 < p < \Delta$. As a result, the right-hand side of the equation acquired an extra contribution $\propto \Delta$. However, this extra term only slightly modifies the solution in the limit of weak disorder. Moreover, it can be absorbed into a redefinition of the parameters $\alpha$ and $\epsilon$ and does not show up in the observable quantities.

After analytic continuation of Eq.\ (\ref{SCBA3D}) in the upper complex half-plane $i\epsilon \mapsto E + i0$, we obtain the retarded self energy
\begin{equation}\label{SigmaR}
 \Sigma^R
  = E + \frac{\sqrt{1 - 2i\pi\alpha E} - 1}{i\pi\alpha}
  \approx -\frac{i\pi\alpha E^2}{2} + \frac{\pi^2\alpha^2 E^3}{2} + \ldots
\end{equation}
The real part of the self energy can be always hidden in the renormalization of the Fermi energy. In turn, the imaginary part is directly observable since it defines the electron scattering rate
\begin{equation}\label{gamma}
 \gamma
  = \frac{1}{2\tau}
  = -\operatorname{Im} \Sigma^R
  \approx \frac{\pi\alpha E^2}{2}.
\end{equation}
We see that our SCBA approach within dimensional regularization scheme correctly reproduces the leading Fermi golden rule estimate of this rate, cf.\ Eq.\ (\ref{Etau}).

\subsection{Density of states}

Average density of states in the presence of disorder is given by Eq.\ (\ref{rhotrace}). This equation contains exactly the same momentum integral as in the right-hand side of the SCBA equation (\ref{SCBA}). This allows us to express the density of states via the self energy directly in 3D:
\begin{equation}
 \rho(E)
  = -\frac{2}{\pi} \operatorname{Im} \int (d^3p)\, \frac{E - \Sigma^R}{(E - \Sigma^R)^2 - p^2}
  = -\frac{\operatorname{Im} \Sigma^R}{\pi^3 \alpha}.
\end{equation}
Using the solution (\ref{SCBA3D}), we obtain a rather simple and closed expression for the density of states
\begin{equation}\label{SCBADOS}
 \rho(E)
  = \frac{\operatorname{Re} \sqrt{1 + 2i\pi\alpha E} - 1}{\pi^4 \alpha^2}
  \approx \frac{E^2}{2\pi^2} \biggl( 1 - \frac{5}{4}\: \pi^2 \alpha^2 E^2 + \ldots \biggr).
\end{equation}
In the limit of weak disorder, the density of states acquires a small negative correction $\propto \alpha^2 E^4$. In Sec.\ \ref{sec:beyond}, we will show that other diagrams, not included in SCBA, provide a stronger correction in this limit. Let us also point out that the SCBA result for the density of states is an analytic function at small $E$. Corrections beyond SCBA will violate this property as well.

\subsection{Conductivity}

\begin{figure}
\centerline{\includegraphics[width = 0.9\columnwidth]{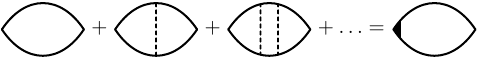}}
\caption{Diagrams for conductivity with the current vertex correction.}
\label{fig:Kubo}
\end{figure}

Semiclassical conductivity is defined by the Kubo formula (\ref{Kubo}) with the two Green functions averaged separately. In addition, the current vertex correction shown in Fig.\ \ref{fig:Kubo} should be included in order to account for all the diagrams with non-intersecting impurity lines. As we will see shortly, each rung of the ladder diagram multiplies the current operator by the same constant $W$:
\begin{equation}\label{W}
 W \sigma_x
  = \alpha \int G^R(i\epsilon_R, \mathbf{p}) \sigma_x G^A(-i\epsilon_A, \mathbf{p})\: p^{d-1}\: dp.
\end{equation}
Here we use Matsubara representation with both energies $\epsilon_{R/A}$ positive, cf.\ Eq.\ (\ref{Matsubara}). These energies also include the corresponding self-energy parts.

Averaging over directions of $\mathbf{p}$ and applying anticommutation rules (\ref{Clifford}), we simplify the integrand in  Eq.\ (\ref{W}) and observe that the right-hand side is indeed proportional to $\sigma_x$ and hence the current operator retains its matrix form. The remaining integral over $p$  defines the factor $W$:
\begin{equation}\label{Wsplit}
 W
  = \alpha \int \frac{\epsilon_R \epsilon_A + p^2 (2-d)/d}{(\epsilon_R^2 + p^2)(\epsilon_A^2 + p^2)}\: p^{d-1}\: dp.
\end{equation}
Here the integrand can be split into two parts with either retarded or advanced denominator. This separation allows us to express the integrals of the two terms through the self energy using the SCBA equation (\ref{SCBA}):
\begin{equation}
 W
  = \frac{\bigl[(d-2)\epsilon_R + d\epsilon_A \bigr] \Sigma^R + \bigl[d \epsilon_R + (d-2)\epsilon_A \bigr] \Sigma^A}
    {i d (\epsilon_R^2 - \epsilon_A^2)} .
\end{equation}
Finally, we perform analytic continuation to real energies according to the rules
\begin{equation}
 i\epsilon_R \mapsto E - \Sigma^R,
 \qquad
 -i\epsilon_A \mapsto E - \Sigma^A
\end{equation}
and get the following result:
\begin{equation}\label{Wresult}
 W
  = \frac{E + (d-2) \operatorname{Re} \Sigma}{d (E - \operatorname{Re} \Sigma)}.
\end{equation}

The diagrams in Fig.\ \ref{fig:Kubo} represent a simple geometric series with the denominator $W$. Summing up this series and setting $d = 3$, we obtain the conductivity
\begin{equation}
 \sigma
  = \frac{1}{\pi^2\alpha} \frac{W}{1 - W}
  = \frac{1}{2\pi^2\alpha} \frac{E + \operatorname{Re} \Sigma}{E - 2\operatorname{Re} \Sigma}.
\end{equation}
With the solution (\ref{SigmaR}), we can get a closed expression for the conductivity including all diagrams with non-intersecting impurity lines.
\begin{multline}\label{Drude}
 \sigma
  = \frac{1}{2\pi^2\alpha}
    \frac{1 - 2\operatorname{Re} \sqrt{1 + 2i\pi\alpha E}}{\operatorname{Re} \sqrt{1 + 2i\pi\alpha E} - 2} \\
  \approx \frac{1}{2\pi^2\alpha} \biggl(
     1 + \frac{3}{2}\: \pi^2 \alpha^2 E^2 + \ldots
    \biggr).
\end{multline}

Exactly at the Weyl point, $E = 0$, the conductivity remains finite. Its value is $\propto 1/\alpha$. This result fully agrees with the previous studies \cite{Ominato14, Klier19}. Equation (\ref{Drude}) suggests a correction $\propto \alpha E^2$ to this constant. In the next Section, we will show that other diagrams, not included in SCBA, provide a stronger correction, that is also non-analytic at small energies.

\section{Interference corrections}
\label{sec:beyond}

In the previous Section, we have calculated the average density of states and conductivity of a Weyl semimetal using the self-consistent Born approximation. This approach automatically includes the complete set of diagrams with non-intersecting impurity lines. In a sense, it is a mean-field approach neglecting possible interference of electrons scattering on different impurities. We will now take into account such interference effects and consider diagrams with crossed impurity lines.

The most prominent effect based on quantum interference of electrons is Anderson localization \cite{Evers08}. It has numerous forms depending on the symmetry of disordered Hamiltonian, on certain topological features of the spectrum, and on the system dimensionality. For the case of 3D Weyl semimetals, as for any other 3D material, localization effects are weak unless disorder strength exceeds a certain threshold value. This weak localization correction in 3D is $\delta\sigma \sim 1/l$, where $l$ is the electron's mean free path. Using Eq.\ (\ref{gamma}), we can estimate the weak localization correction in our model as $\delta\sigma \sim \alpha E^2$. Such a correction is the result of summation of an infinite set of maximally crossed diagrams \cite{Gorkov79}. Quite curiously, weak localization effect is of the same order as the correction to conductivity due to non-intersecting diagrams (\ref{Drude}). In this Section we will calculate crossed diagrams with two and three impurity lines. It will be shown that these diagrams provide a stronger interference correction than the weak localization effect.

Our calculation is technically similar to the treatment of ballistic interference effects in a conventional disordered metal with parabolic dispersion. Such a calculation was carried out in Refs.\ \cite{Kirk86, Wysokinski95}. However, in the case of Weyl semimetals additional complications arise due to the matrix structure of the Hamiltonian (\ref{ham}).

\subsection{Density of states}

\begin{figure}
\centerline{\includegraphics[width = 0.9\columnwidth]{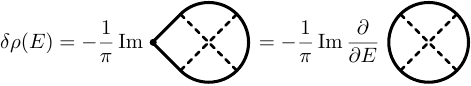}}
\caption{Interference correction to the density of states in two equivalent forms: as a Green function at coincident points and as an energy derivative of a vacuum diagram.}
\label{fig:DOScrossed}
\end{figure}

The first non-trivial diagram that provides an interference correction to the density of states (\ref{SCBADOS}) involves two crossed impurity lines. This diagram is shown in Fig.\ \ref{fig:DOScrossed}. Since all the Green functions in this diagram are taken at the same energy, we can considerably simplify the calculation by computing the corresponding vacuum diagram first and then taking its derivative in energy:
\begin{gather}
 \delta\rho(E)
  = -\frac{1}{\pi} \operatorname{Im} \frac{\partial F_2(E + i0)}{\partial E}, \label{drho} \\
 F_2(i\epsilon)
  = -\frac{\alpha^2}{4} \int (d^d q) \operatorname{tr} \Bigl[
      \Pi(i\epsilon, \mathbf{q}) \Pi(i\epsilon, -\mathbf{q})
    \Bigr], \label{F2} \\
 \Pi(i\epsilon, \mathbf{q})
  = \int p^{d-1}\, dp\, \frac{
      (i\epsilon + \bm{\sigma}\cdot \mathbf{p} + \bm{\sigma}\cdot \mathbf{q})(i\epsilon + \bm{\sigma}\cdot \mathbf{p})
    }{[\epsilon^2 + (\mathbf{p + q})^2][\epsilon^2 + p^2]}
    \label{Pi}.
\end{gather}
Strictly speaking, we should retain the self-energy contribution in the Green functions and perform analytic continuation from a Matsubara energy $i\epsilon$ to the energy $E + i\gamma$. However, for the calculation of the density of states, the extra imaginary part of the self energy $\gamma$ can be neglected since it provides a correction of a higher order in $\alpha E$.

Four Green functions in the vacuum diagram Fig.\ \ref{fig:DOScrossed} are split into two similar pairs and the result of momentum integration inside each pair is denoted by $\Pi$. Let us first analyze this latter integral of the product of two Green functions. Let us split the momentum $\mathbf{p} = \mathbf{p}_\parallel + \mathbf{p}_\perp$ into components along and perpendicular to the vector $\mathbf{q}$. The denominator of the integrand in Eq.\ (\ref{Pi}) does not depend on the direction of $\mathbf{p}_\perp$ hence we can average the numerator with respect to this direction. Effectively, it means dropping all the terms which are odd in $\mathbf{p}_\perp$. Next, we can express the parallel component through the scalar product $\mathbf{pq}$ and replace the latter with the help of the identity $2\mathbf{pq}  = (\mathbf{p+q})^2 - p^2 - q^2$. Then we split the integrand into separate fractions such that their numerators do not contain $\mathbf{p}$. This yields the following expression:
\begin{multline}
 \Pi(i\epsilon, \mathbf{q})
  = \int p^{d-1}\, dp\, \Biggl[
      \frac{-2\epsilon^2 - q^2/2}{[\epsilon^2 + (\mathbf{p + q})^2][\epsilon^2 + p^2]} \\
      + \frac{1}{\epsilon^2 + p^2}
    \Biggr].
\end{multline}
Here we have also shifted the integration variable $\mathbf{p} \mapsto \mathbf{p - q}$ in one of the terms of the integrand. Note that after all the transformations, the integral $\Pi$ acquired a trivial matrix structure.

We have represented $\Pi$ as a combination of two basic integrals:
\begin{subequations}\label{II}
\begin{align}
 I(i\epsilon)
  &= \int \frac{p^{d-1}\, dp}{\epsilon^2 + p^2}
  = \frac{\pi \epsilon^{d-2}}{2 \sin(\pi d/2)}, \label{IIa}\\
 I(i\epsilon, i\epsilon, q)
  &= \int \frac{p^{d-1}\, dp}{[\epsilon^2 + (\mathbf{p + q})^2][\epsilon^2 + p^2]}.
\end{align}
\end{subequations}
The first of these integrals we have already encountered earlier in the calculation of the self energy (\ref{SCBA}). The second integral is more complicated since its denominator has a nontrivial dependence on the angle between $\mathbf{p}$ and $\mathbf{q}$. This latter integral is in fact similar to the polarization operator for a Hamiltonian with quadratic dispersion. Calculation of such a polarization operator and its analytic continuation to real energies is outlined in Appendix \ref{App}. The final result for $\Pi$ is
\begin{equation}
 \Pi(i\epsilon, \mathbf{q})
  = I(i \epsilon) - \left( 2\epsilon^2 + \frac{q^2}{2} \right) I(i \epsilon, i \epsilon, q).
\end{equation}

Once the value of $\Pi$ is known, we can readily use Eq.\ (\ref{F2}) and represent the vacuum diagram as a single integral in $q$. At the same time, we perform analytic continuation to real energies and obtain
\begin{equation}\label{F2int}
 F_2(E + i0)
  = -\frac{\alpha^2}{2} \int (d^d q) \left[
      I_R + \left( 2E^2 - \frac{q^2}{2} \right) I_{RR}(q)
    \right]^2.
\end{equation}
Here $I_R$ is the retarded form of the integral Eq.\ (\ref{IIa}) and $I_{RR}$ is the retarded-retarded form of the polarization operator (\ref{IRRAA}).

Integral (\ref{F2int}) diverges at large $q$ in 3D. We can extract this divergent part by using the simplified form of the polarization operator (\ref{IRRarctan}) and expanding in $q \gg E$:
\begin{multline}\label{F2UV}
 F_2(E + i0)
  = \text{const} - \alpha^2 \int_E^\infty dq \Big[
      \frac{\pi^2}{256}\: q^{3d-9} (q^2 - 4E^2)^2\\
      - \frac{i \pi}{60}\: E^d q^{2d-7} (5q^2 - 16E^2)
      \Big].
\end{multline}
Lower bound of the integral is set to $E$ since this is effectively the only available parameter of the proper dimension. The exact value of this lower bound is immaterial since we can always change it at the cost of redefining the constant term.

Dimensional regularization implies that the divergent integral in Eq.\ (\ref{F2UV}) is calculated assuming $d$ is low enough ($d < 4/3$ in this case) and then the result is analytically continued to $d = 3$. Strongly divergent terms provide some finite contributions after this procedure. This is fully analogous to how the divergent real part of the self energy is removed by renormalization of the chemical potential. The only non-trivial contribution comes from the very last term whose divergence in 3D is logarithmic.
\begin{multline}\label{F2pole}
 F_2(E + i0)
  = \text{const} - \frac{4 i \pi \alpha^2}{15}\: E^{d+2} \int_E^\infty dq\, q^{2d-7}\\
  = \text{const} + \frac{2 i \pi \alpha^2}{15}\: \frac{E^{3d-4}}{d-3}.
\end{multline}
Thus, the function $F_2$ acquires a simple pole at $d = 3$. This is a standard manifestation of a logarithmic divergence in the dimensional scheme. We can convert this pole into explicit logarithm by expanding the numerator near $d = 3$:
\begin{multline}
 F_2(E + i0)
  = \frac{2 i \pi}{15}\: \alpha^2 E^5 \left( \frac{1}{d-3} + 3 \ln E + \ldots \right) \\
  = \frac{2 i \pi}{5}\: \alpha^2 E^5 \ln|E/\Delta|.
\end{multline}
The divergent term $1/(d-3)$ should be replaced by some large number related to the ultraviolet cutoff scale $\Delta$. We simply incorporate this scale into the logarithmic factor as shown in the last expression.

Interference correction to the average density of states can be found from Eq.\ (\ref{drho}). Together with the SCBA expression (\ref{SCBADOS}), this gives the final result
\begin{equation}\label{DOS}
 \rho(E)
  = \frac{E^2}{2\pi^2} - 2 \alpha^2 E^4 \ln|E/\Delta| + O(\alpha^2 E^4).
\end{equation}
Remarkably, interference correction to the density of states is stronger than a similar correction from SCBA due to just an extra logarithm factor. At the same time, this logarithmic correction is non-analytic as a function of energy in the limit $E \to 0$ unlike the SCBA result.

\subsection{Conductivity}

Let us now consider similar interference corrections to the quasiclassical conductivity (\ref{Drude}). The calculation will be carried out exactly with the same strategy as we did for the density of states. First, we will reduce every diagram to a single $d$-dimensional integral over momentum $q$ in terms of polarization operators. Then we will expand the integrand in the small ratio $\gamma/E$ and retain only relevant leading terms of this expansion. Finally, we will analyse ultraviolet behavior of each term and apply dimensional scheme to treat divergent contributions.

\subsubsection{Diagrams with two crossed impurity lines}

\begin{figure}
\centerline{\includegraphics[width = 0.9\columnwidth]{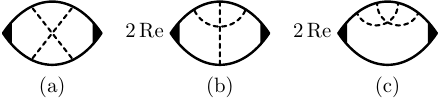}}
\caption{\label{fig:2lines} Interference corrections to conductivity with two crossed impurity lines.}
\end{figure}

There are in total three different interference diagrams for conductivity with two crossed impurity lines, see Fig.\ \ref{fig:2lines}. All these diagrams can be generated from the single vacuum diagram Fig.\ \ref{fig:DOScrossed} by inserting two current vertices in different positions. However, unlike the case of the density of states, we cannot calculate conductivity by taking derivatives of the vacuum diagram in some parameter because the conductivity diagram involves both retarded and advanced Green functions.

Consider the diagram Fig.\ \ref{fig:2lines}a.
\begin{multline}
 \delta\sigma_{2a}
  = \frac{\alpha^2}{(1-W)^2} \int (d^d q)\: p_1^{d-1}\, dp_1\: p_2^{d-1}\, dp_2 \\
    \times \operatorname{tr} \Bigl[
      \sigma_x G^R(\mathbf{p}_1 + \mathbf{q}) G^R(\mathbf{p}_2 + \mathbf{q}) G^R(-\mathbf{p}_1) \\
      \times \sigma_x G^A(-\mathbf{p}_1) G^A(-\mathbf{p}_2) G^A(\mathbf{p}_1 + \mathbf{q})
    \Bigr].
\end{multline}
Our first goal is to reduce the integrand to a product of polarization operators. Each polarization operator will incorporate integration over $p_1$ or $p_2$ and the $q$ integral will be analyzed later. Our expression contains four Green functions which involve the momentum $p_1$. Therefore we first split the integrand into fractions involving only two factors with $\mathbf{p}_1$ in the denominator. This is fully analogous to the splitting of denominators in Eq.\ (\ref{Wsplit}). The next step is taking the trace of $\sigma$ matrices in the numerator and averaging over directions of $\mathbf{p}_{1,2}$ as was explained earlier in the calculation of the density of states. Finally, we split the integrand in individual fractions whose numerators are independent of $\mathbf{p}_{1,2}$. This way we represent $\delta\sigma_{2a}$ as a single $q$ integral of a quadratic expression in terms of polarization operators.
\begin{equation}\label{dsa}
 \delta\sigma_{2a}
  = \int_0^\infty dq\: S_{2a}(q).
\end{equation}

\begin{figure}
\centerline{\includegraphics[scale=0.85]{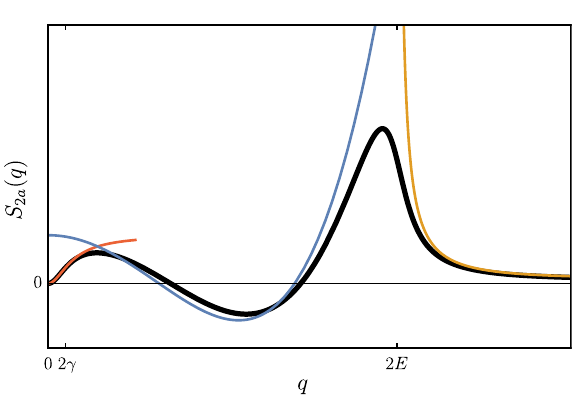}}
\caption{Momentum dependence of the integrand $S_{2a}(q)$ for the representation (\ref{dsa}) of the diagram Fig.\ \ref{fig:2lines}a (thick black curve). Parameters of the plot $\gamma/E = 0.05$. Three asymptotic expressions Eqs.\ (\ref{saq}), (\ref{saq123}) are shown by the red, blue, and orange lines, respectively.}
\label{fig:Xintegrand}
\end{figure}

Explicit form of the function $S_{2a}(q)$ for an arbitrary $d$ is very cumbersome hence we only show the integrand graphically in Fig.\ \ref{fig:Xintegrand} for the case $d = 3$. There are three qualitatively different regions in this function:
\begin{equation}\label{saq}
 S_{2a}(q)
  = \begin{cases}
      S^\text{I}_{2a}(q), & q \lesssim \gamma, \\
      S^\text{II}_{2a}(q), & \gamma \ll q < 2E, \\
      S^\text{III}_{2a}(q), & q > 2E.
    \end{cases}
\end{equation}
To analyse the integral, we can do further simplifications by setting $d = 3$ in the first two regions and by expanding to the leading orders in the small parameter $\gamma/E$. In the third region, we have also expanded in $q \gg E$ and retained only the relevant leading terms. This way we obtain the following expressions:
\begin{subequations}
\label{saq123}
\begin{align}
 S^\text{I}_{2a}(q)
  &= -\frac{3}{8\pi^2} \left[
      \frac{2\gamma}{q} - \left( 1 + \frac{4\gamma^2}{q^2} \right) \arctan\frac{q}{2\gamma}
    \right] \notag \\ & \qquad\quad \times \left[
      \frac{6\gamma}{q} + \left( 1 - \frac{12\gamma^2}{q^2} \right) \arctan\frac{q}{2\gamma}
    \right], \label{sa1q} \\
 S^\text{II}_{2a}(q)
  &= \frac{9}{32} \left(
      \frac{1}{3} - \frac{q^2}{E^2} + \frac{7q^4}{16 E^4} - \frac{q^6}{96 E^6}
    \right), \label{sa2q} \\
 S^\text{III}_{2a}(q)
  &= \frac{3 \alpha}{32}\: q^{2 d-5} \left[
      \frac{2}{3}+\frac{8 E^2}{5 q^2} + O\bigl( q^{-4} \bigr)
    \right]. \label{sa3q}
\end{align}
\end{subequations}
These three asymptotic forms are also illustrated in Fig.\ \ref{fig:Xintegrand}. Let us point out that parameters $\gamma$ and $\alpha$ are related by Eq.\ (\ref{gamma}) in 3D. For $d \approx 3$, a slightly more general relation
\begin{equation}
 \gamma = \frac{\pi}{2}\: \alpha E^{d-1}
\end{equation}
should be used.

The first asymptotic region $q \lesssim \gamma$ provides a contribution to the conductivity $\propto \alpha E^2$. This part is normally taken into account in the weak localization correction. We will neglect it in favor of other, larger contributions. Integral over the second region $\gamma \ll q < 2E$ converges and provides a correction \cite{Gorkov79}
\begin{equation}\label{Xleading}
 \int_0^{2E} dq\: S^\text{II}_{2a}(q)
  = \frac{6 |E|}{35}.
\end{equation}
This correction is smaller than the leading Drude result (\ref{Drude}) but, at the same time, larger than both weak localization and the subleading term in Eq.\ (\ref{Drude}). Remarkably, unlike the SCBA result, this term is of an odd power in $E$ and hence exhibits a cusp at $E = 0$. We stress this fact by writing explicitly the absolute value $|E|$ in Eq.\ (\ref{Xleading}) thus allowing for $E$ of any sign.

Contribution of the third asymptotic region $q > 2E$ has an extra $\alpha$ factor compared to $S^\text{II}_{2a}$ hence it should be negligible at first sight. However, it contains an ultraviolet divergence that can lead to an extra logarithmic factor as we have already seen in the calculation of the density of states. In the expression (\ref{sa3q}), we have retained exactly the terms that provide such an ultraviolet divergence for $d = 3$.
\begin{multline}
 \int_{2E}^\infty dq\: S^\text{III}_{2a}(q)
  = \text{const} + \alpha \int_{2E}^\infty dq \;\frac{3 E^2 q^{2d - 7}}{20} \\
  = \text{const} - \frac{3 \alpha E^{2d-4}}{40 (d-3)}
  = \text{const} - \frac{3}{20}\: \alpha E^2 \ln|E/\Delta|.
\end{multline}
We conclude that the contribution of this part, although having an extra $\alpha E$ factor as compared to Eq.\ (\ref{Xleading}), is still dominant compared to both the weak localization and the SCBA subleading term (\ref{Drude}) due to an extra logarithm. Collecting all the terms together, we have the following result for the diagram Fig. \ref{fig:2lines}a:
\begin{subequations}
\label{s2}
\begin{equation}
 \delta\sigma_{2a}
  = \frac{6 |E|}{35} - \frac{3}{20}\: \alpha E^2 \ln|E/\Delta|.
\end{equation}

Calculation of the other two diagrams in Fig.\ \ref{fig:2lines} is performed in exactly the same way. The only slight technical difference for the diagram Fig.\ \ref{fig:2lines}c is the presence of two identical Green functions. This means that in some terms we will encounter a square of the Green function's denominator. These terms should be expressed via an energy derivative of the polarization operator. The results for the two diagrams are
\begin{align}
 \delta\sigma_{2b}
  &= \frac{3}{10}\: \alpha E^2 \ln|E/\Delta|, \\
 \delta\sigma_{2c}
  &= -\frac{2 |E|}{5} + \frac{1}{4}\: \alpha E^2 \ln|E/\Delta|.
\end{align}
\end{subequations}

Overall interference correction to the conductivity from diagrams Fig.\ \ref{fig:2lines} is the sum of Eqs.\ (\ref{s2}):
\begin{equation}\label{deltas2}
 \delta\sigma_2
  = -\frac{8 |E|}{35} + \frac{2}{5}\: \alpha E^2 \ln|E/\Delta|.
\end{equation}
This result contains both the leading ($\propto \alpha^0$) and the subleading ($\propto \alpha^1$) terms. Higher diagrams with three crossed lines have an extra $\alpha$ factor and can provide additional contributions to the subleading interference correction only. These diagrams will be studied in the next Section.

\subsubsection{Diagrams with three crossed impurity lines}

\begin{figure}
\centerline{\includegraphics[width = 0.9\columnwidth]{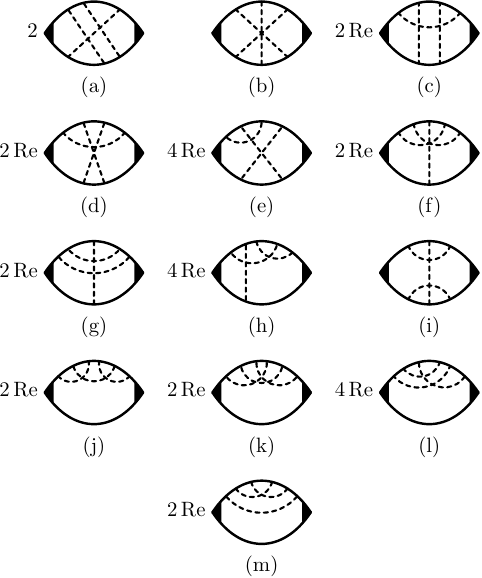}}
\caption{Interference corrections to the conductivity with three crossed impurity lines.}
\label{fig:3lines}
\end{figure}

We will now consider interference diagrams for conductivity with three crossed impurity lines. They are shown in Fig.\ \ref{fig:3lines}. All these diagrams except the last one can be generated from just two distinct vacuum diagrams by inserting two current vertices in all possible positions. Each of the two corresponding vacuum diagrams can be represented as integrals over a single momentum $q$ of a product of three polarization operators. This property is preserved after insertion of current vertices since they can only double certain denominators, that are later decoupled. Hence we can calculate all but the last three-impurity diagrams with the same method as was described in the previous section for diagrams with two impurities. As a typical example, we will explain the calculation of the first diagram Fig.\ \ref{fig:3lines}a and then give the results for all the other diagrams.

\begin{figure}
\centerline{\includegraphics[scale=0.85]{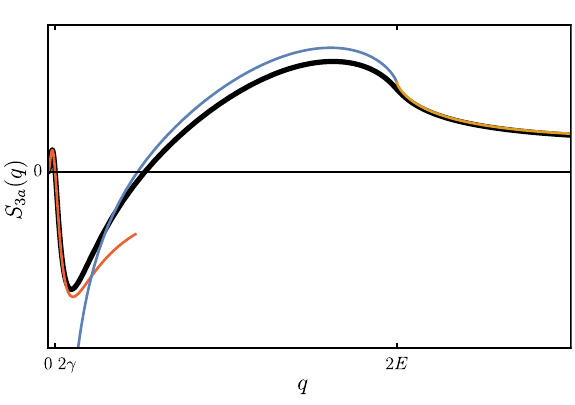}}
\caption{Momentum dependence of the integrand $S_{3a}(q)$ for the representation (\ref{dsd}) of the diagram Fig.\ \ref{fig:3lines}a (thick black curve). Parameters of the plot $\gamma/E = 0.02$. Three asymptotic expressions (\ref{sdq}) are shown by the red, blue, and orange lines.}
\label{fig:Dintegrand}
\end{figure}

The diagram Fig.\ \ref{fig:3lines}a can be written as a momentum integral of up to three polarization operators (\ref{Iarctan}) after taking the trace of all the Green functions, averaging over directions of momenta $\mathbf{p}_{1,2,3}$, and algebraically splitting into individual fractions whose numerators are independent of $\mathbf{p}_{1,2,3}$. All these steps are completely analogous to the calculation of the two-impurity diagrams in the previous Section. After extensive algebra, we arrive at the single $q$ integral
\begin{equation}\label{dsd}
 \delta\sigma_{3a}
  = \int_0^\infty dq\, S_{3a}(q).
\end{equation}
The integrand of this expression is shown in Fig.\ \ref{fig:Dintegrand}. We can again approximate it in three asymptotic regions of small, intermediate, and large $q$:
\begin{equation}\label{sdq}
 S_{3a}(q)
  = \begin{cases}
      S^\text{I}_{3a}(q), & q \lesssim \gamma, \\
      S^\text{II}_{3a}(q), & \gamma \ll q < 2E, \\
      S^\text{III}_{3a}(q), & q > 2E.
    \end{cases}
\end{equation}
These asymptotic forms are calculated in the same way as we did it earlier for the two-impurity diagram. They are also shown in Fig.\ \ref{fig:Dintegrand}, however, explicit asymptotic expressions are still too bulky to write them here explicitly.

The first region provides a finite correction of the order $\alpha E^2$ completely similar to the previously studied case of two impurities. We will disregard this contribution altogether. In the second region we now have one extra $\alpha$ factor in comparison to the diagram with two impurities. Hence it seems that we can neglect the contribution from this region as well. However, as can be seen from Fig.\ \ref{fig:Dintegrand}, this intermediate region develops a new type of divergence towards lower values of $q$. Specifically, we can extract the following asymptotics:
\begin{equation}
 S_{3a}(\gamma \ll q \ll E)
  = -\frac{3 (16 - \pi^2)}{64}\: \frac{\alpha E^2}{q}.
\end{equation}
Integral of this term is not truly divergent since it is limited by $\gamma$ from below. But it does provide a logarithmic correction:
\begin{equation}
 \int_0^{2E} dq\, S_{3a}(q)
  = \text{const} + \frac{3 (16 - \pi^2)}{64}\: \alpha E^2 \ln|\alpha E|.
\end{equation}
The constant term here is of the order $\alpha E^2$ without any logarithm.

Finally, we can evaluate the contribution of the region $q > 2E$. Full asymptotic expression $S^\text{III}_{3a}$ is also not needed for this calculation. It suffices to take only divergent terms in the limit $q \to \infty$ and apply the dimensional regularization recipe.
\begin{multline}
 \int_{2E}^\infty dq\, S_{3a}(q)
  = \text{const} + \alpha \int_{2E}^\infty dq \;\frac{23 E^2 q^{2d - 7}}{60} \\
  = \text{const} - \frac{23 \alpha E^{2d-4}}{120 (d-3)}
  = \text{const} - \frac{23}{60}\: \alpha E^2 \ln|E/\Delta|.
\end{multline}
Full contribution of the diagram Fig.\ \ref{fig:3lines}a is thus
\begin{subequations}
\label{s3}
\begin{equation}
 \delta \sigma_{3a}
  = \alpha E^2 \left[ \frac{3(16 - \pi^2)}{64}\, \ln|\alpha E| - \frac{23}{60}\, \ln|E/\Delta| \right].
\end{equation}

All other diagrams of Fig.\ \ref{fig:3lines} except (m) are evaluated in the same way and the results are
\begin{align}
 \delta \sigma_{3b}
  &= \frac{3(16 - 3 \pi^2)}{64}\, \alpha E^2 \ln|\alpha E|, \\
 \delta \sigma_{3c}
  &= \alpha E^2 \left[ \frac{3(\pi^2 - 16)}{64} \ln|\alpha E| - \frac{41}{60}\, \ln|E/\Delta| \right], \\
 \delta \sigma_{3d}
  &= \alpha E^2 \left[ \frac{9\pi^2 + 16}{384}\, \ln|\alpha E| + \frac{1}{60}\, \ln|E/\Delta| \right], \\
 \delta \sigma_{3e}
  &= \frac{13}{30}\, \alpha E^2 \ln|E/\Delta|, \\
 \delta \sigma_{3f}
  &= -\frac{7}{12}\, \alpha E^2 \ln|E/\Delta|, \\
 \delta \sigma_{3g}
  &= 0, \\
 \delta \sigma_{3h}
  &= -\frac{1}{6}\, \alpha E^2 \ln|E/\Delta|, \\
 \delta \sigma_{3i}
  &= 0, \\
 \delta \sigma_{3j}
  &= -\frac{2}{3}\, \alpha E^2 \ln|E/\Delta|, \\
 \delta \sigma_{3k}
  &= \frac{47}{12}\, \alpha E^2 \ln|E/\Delta|, \\
 \delta \sigma_{3l}
  &= -\frac{4}{3}\, \alpha E^2 \ln|E/\Delta|.
\end{align}

It remains to calculate the last diagram Fig.\ \ref{fig:3lines}m. Not all impurity lines intersect in this diagram hence it decouples into two separate integrals. Consider first the part with two current vertices and three adjacent Green functions. We observe that after $\mathbf{p}$ integration this part of the diagram has a trivial matrix structure and can be taken out of the overall matrix trace. Then, the rest of the diagram with two crossed impurity lines is identical to the diagram Fig.\ \ref{fig:DOScrossed} for the interference correction to the density of states. Using Eq.\ (\ref{F2pole}), we obtain the following result for the correction to conductivity:
\begin{equation}
 \delta \sigma_{3m}
  = \frac{3 E^{1-d}}{2\pi \alpha} \operatorname{Im} \frac{\partial F_2}{\partial E}
  = 2 \alpha E^2 \ln|E/\Delta|.
\end{equation}
\end{subequations}

Summing up all three-impurity diagrams (\ref{s3}), we obtain
\begin{equation}
 \delta\sigma_3
  = \left( \frac{5}{12} - \frac{3\pi^2}{64} \right) \alpha E^2 \ln|\alpha E|
    +\frac{51}{20}\, \alpha E^2 \ln|E/\Delta|.
\end{equation}
Together with the SCBA expression (\ref{Drude}) and with the corrections from diagrams with two impurities (\ref{deltas2}), we have the final result for conductivity
\begin{multline}\label{sigma23}
 \sigma
  = \frac{1}{2\pi^2 \alpha} - \frac{8 |E|}{35}
    +\left( \frac{5}{12} - \frac{3\pi^2}{64} \right) \alpha E^2 \ln|\alpha E|\\
    +\frac{59}{20}\, \alpha E^2 \ln|E/\Delta| + O(\alpha E^2).
\end{multline}
We have thus  established the leading and the subleading terms in the energy dependence of conductivity.

The result (\ref{sigma23}) is written in terms of the disorder strength parameter $\alpha$ and the Fermi energy $E$. Both these parameters also include some uncontrollable renormalization constants that were implicitly included by the dimensional regularization procedure [e.g.\ an ultraviolet divergent part of the self energy in Eq.\ (\ref{SCBA})]. Neither the Fermi energy nor the disorder strength are observable parameters of the material. At the same time, conductivity can be measured directly as a function of electron concentration. The latter can be also controlled, at least in principle, by chemical doping or external gating.

Electron concentration is related to the Fermi energy by integrating the density of states. For our purpose, it is in fact sufficient to retain just the leading term in Eq.\ (\ref{DOS}), which is the density of states of a clean Weyl semimetal (\ref{rho0})
\begin{equation}
 n(E)
  = \int_0^E dE\, \rho(E)
  = \frac{E^3}{6\pi^2},
 \qquad
 E
  = (6 \pi^2 n)^{1/3}.
\end{equation}
We substitute this expression for energy into Eq.\ (\ref{sigma23}) and find
\begin{equation}\label{sigman}
 \sigma
  =  \sigma_0 + A n^{1/3}
    + \frac{n^{2/3}}{\sigma_0} \Bigl( B \ln|n/\sigma_0^3| + C \ln|n/\Delta^3| \Bigr).
\end{equation}
We have also introduced here the parameter $\sigma_0$ corresponding to the conductivity at zero doping (in units $e^2/h$). This parameter is also directly measurable and replaces $\alpha$. Our result (\ref{sigman}) is thus a relation involving only observable quantities. Three constants in this relation are
\begin{align}
 A &= -\frac{8}{35} (6 \pi^2)^{1/3} \approx -0.8909, \\
 B &= \frac{(6 \pi^2)^{2/3}}{16 \pi} \left(\frac{5}{3\pi^2}-\frac{3}{16}\right) \approx -0.0056, \\
 C &= \frac{59}{40 \pi (6 \pi^2)^{1/3}} \approx 0.1205.
\end{align}

While the electron concentration is at least partially controllable in the experiment, it seems unfeasible that amount of disorder can be changed at will. That is why the two logarithmic terms in Eq.\ (\ref{sigman}) can hardly be distinguished. A simplified version of our result with the two logarithms combined reads
\begin{equation}
 \sigma
  = \sigma_0 + A n^{1/3} + (B + C) \frac{n^{2/3}}{\sigma_0} \ln|n|.
\end{equation}
Here a normalization constant under the logarithm is unspecified and the prefactor of the second term is $B + C \approx 0.1148$.

\section{Summary and discussion}
\label{sec:conclusion}

To summarize, we have studied spectral and transport properties of a 3D Weyl semimetal in the presence of Gaussian white-noise disorder. We have developed a perturbation theory approach in the weak disorder limit controlled by the parameter $\alpha E \ll 1$ using dimensional regularization scheme near $d = 3$. Both semiclassical and interference contributions were taken into account to calculate the average density of states and conductivity.

In the framework of the self-consistent Born approximation, we have found a closed analytic expression for the self energy in three dimensions (\ref{SCBA3D}). This allowed us to derive full mean-field results for the density of states (\ref{SCBADOS}) and conductivity (\ref{Drude}) taking into account the whole set of non-intersecting diagrams. Both quantities are analytic functions in the limit $E \to 0$ and have a regular expansion in powers of the small parameter $\alpha^2 E^2$.

We have also considered interference corrections due to diagrams with two and three intersecting impurity lines. These interference terms are dominant in comparison with the semiclassical result and also non-analytic in the limit $E \to 0$. Density of states includes an extra ultraviolet logarithmic correction (\ref{DOS}) as compared to the mean-field result (\ref{SCBADOS}). Conductivity acquires a leading universal (independent of the disorder strength) correction $\propto |E|$ and two subleading corrections with the ultraviolet and infrared logarithms (\ref{sigma23}). The leading correction to the conductivity is due to the diagrams with two crossed impurity lines while the subleading logarihmic terms also include a contribution from the diagrams with three crossed impurities. Our results produce a prediction for the concentration dependence of conductivity (\ref{sigman}) that can be directly checked in an experiment.

It should be pointed out that our results are obtained at relatively low energies $E \ll \Delta \ll 1/\alpha$. Possible nonperturbative effects disregarded in this work may become important in a yet much smaller range of energies $E \lesssim \Delta \exp[-(\alpha \Delta)^{-1}]$. While the very existence of these nonperturbative contributions is still debated \cite{Sbierski14, Trescher17, DasSarma16-2, Bera16, Roy18,Syzranov15-1, Syzranov15-2, Roy16, Syzranov16, Louvet16, Erratum, Altland15, Nandishore14, Holder17, DasSarma16-1, Buchhold18, Ominato14, Klier19, Pix21}, we would like to stress that our results do apply in a relatively broad range of parameters possibly except an exponentially small region in the vicinity of zero concentration/energy.

It is worth mentioning some similarities between our results and interference corrections in a conventional metal with parabolic spectrum studied in Refs.\ \cite{Kirk86, Wysokinski95}. The leading correction in both models is of a relative strength $\alpha E \sim (E \tau)^{-1}$ and comes from the same two diagrams Fig.\ \ref{fig:2lines}(a,c). The subleading correction with the infrared logarithm has a relative strength $(\alpha E)^2 \ln(\alpha E) \sim (E \tau)^{-2} \ln(E\tau)$. In a conventional metal it comes from all three diagrams with two crossed impurities Fig.\ \ref{fig:2lines} and in addition from four diagrams Fig.\ \ref{fig:3lines}(a--d). In a Weyl semimetal only the latter four diagrams provide this correction. Finally, the ultraviolet logarithmic correction $\sim (\alpha E)^2 \ln(E/\Delta)$ comes from almost all interference diagrams with two and three impurities considered in this work, while in the case of conventional metal such a correction appears only in two diagrams Fig.\ \ref{fig:2lines}c and Fig.\ \ref{fig:3lines}m. Ultraviolet logarithm cancels in the expression for electron mobility $\mu = \sigma/(en)$ in a conventional metal model while for the Weyl semimetal such a cancellation does not occur.

The model of the random potential disorder can be directly generalized to include a possible random vector potential. This will lead to appearance of two distinct disorder parameters instead of a single quantity $\alpha$. The whole calculation scheme developed in this work can be applied to this more general model with minimal modifications. While the number of relevant diagrams will increase dramatically (each impurity line will be of either scalar or vector type), we expect that qualitative results of our calculation will still hold. Only the coefficients in the expansion (\ref{sigma23}) will be modified. This more general model will be the subject of a separate study.

\begin{acknowledgments}
We are grateful to I.\ S.\ Burmistrov, I.\ V.\ Gornyi, E.\ J.\ K\"onig, A.\ V.\ Lunkin and B.\ Sbierski for stimulating discussions.
\end{acknowledgments}

\appendix
\section{\label{App} Polarization operator}

In this Appendix, we discuss the momentum integral which involves denominators of two Green functions in an arbitrary dimension $d$. By analogy, we name such an integral polarization operator. In the Matsubara formalism, we have the following expression for the polarization operator:
\begin{equation}\label{Idef}
 I(i\epsilon_1, i\epsilon_2, q)
  = \int \frac{p^{d-1}\, dp}{[\epsilon_1^2 + (\mathbf{p+q})^2][\epsilon_2^2 + p^2]}.
\end{equation}
Energy parameters $\epsilon_{1,2}$ are assumed real and positive. Calculation of the polarization operator starts with the standard Feynmann trick that allows us to combine the two denominators into a single one at the cost of introducing an auxiliary integral. After that, $d$-dimensional integration over momentum is straightforward. Subsequent integration over the Feynman parameter yields a linear combination of hypergeometric functions
\begin{equation} \label{IMM}
 I(i\epsilon_1, i\epsilon_2, q)
  = \frac{\pi \bigl[ M(i\epsilon_1, i\epsilon_2, q) + M(i\epsilon_2, i\epsilon_1, q) \bigr]}{2\sin(\pi d/2)},
\end{equation}
with
\begin{multline}\label{M2}
 M(i \epsilon_1, i \epsilon_2, q) \\
  = \frac{\epsilon_1^{d-2}}{q^2 - \epsilon_1^2 + \epsilon_2^2} F\left(
        1, \frac{1}{2}, \frac{d}{2}, -\frac{4 \epsilon_1^2 q^2}{(q^2 - \epsilon_1^2 + \epsilon_2^2)^2}
      \right) \\
      -\frac{\Gamma^2(d/2)}{\Gamma(d-1)}
      \frac{q^{2-d} \operatorname{sgn} \bigl( q^2 - \epsilon_1^2 + \epsilon_2^2 \bigr)}
      {\bigl[ (q^2 + \epsilon_1^2 + \epsilon_2^2)^2 - 4 \epsilon_1^2 \epsilon_2^2 \bigr]^{(3-d)/2}}.
\end{multline}

Our next goal is to perform analytic continuation from Matsubara energies $\epsilon_{1,2}$ to real energies. Let us begin with the simpler case of a polarization operator involving two Green functions of the same kind. Its calculation amounts to setting $\epsilon_1 = \epsilon_2 = \epsilon$ and performing analytic continuation in the upper complex half-plane $i\epsilon \mapsto i\gamma \pm E$. (Here we include the real part of the self-energy into $E$ while keeping the imaginary part $\gamma$ explicit.) This yields the following result:
\begin{multline} \label{IRRAA}
 I_{RR/AA}(q) \\
  = \frac{\pi}{\sin(\pi d/2)} \Biggl[
      \frac{(\gamma \mp i E)^{d-2}}{q^2} F\left(
        1, \frac{1}{2}, \frac{d}{2}, \frac{4 (E \pm i\gamma)^2}{q^2}
      \right) \\
      -\frac{\Gamma^2(d/2)}{q\, \Gamma(d-1)} \bigl[ q^2 - 4 (E \pm i\gamma)^2 \bigr]^{(d-3)/2}
    \Biggr].
\end{multline}
The case of mixed retarded-advanced polarization operator is more subtle. Analytic continuation in two different energy parameters can be performed as follows. We set $i\epsilon_{1,2} = i\gamma \pm t$ and continuously change $t$ from $0$ to $E$. At the initial value $t = 0$, both energies $\epsilon_{1,2}$ are positive and equal hence the sign function in Eq.\ (\ref{M2}) equals $1$. For relatively low energies $E < q/2$, we can analytically continue $t$ directly along the real axis and reach the point $t = E$ without encountering any singularities. Substituting the end point of this trajectory into Eq.\ (\ref{M2}) yields the result for $M$ and hence for $I_{RA}$.

For larger energies, $E > q/2$, the argument of the hypergeometric function in Eq.\ (\ref{M2}) hits the value $1$ when $t$ reaches $q/2$. This is a branch point of the hypergeometric function. We can circumvent this point in the complex plane of $t$ either above or below the real axis. In both cases, one of the two hypergeometric functions in Eq.\ (\ref{IMM}) stays on its main branch while the argument of the other hypergeometric function crosses the branch cut. We account for this crossing by subtracting the corresponding jump of the hypergeometric function. Effectively, this reverses the sign function in Eq.\ (\ref{M2}) for one of the two terms in Eq.\ (\ref{IMM}) and both sign functions cancel each other. Final result for the mixed retarded-advanced polarization operator is
\begin{multline}\label{IRA}
 I_{RA}(q)
  = \frac{\pi}{2\sin(\pi d/2)} \Biggl[
      \Biggl(
        \frac{(\gamma - i E)^{d-2}}{q^2 + 4 i E \gamma}\\
        \times F\left(
          1, \frac{1}{2}, \frac{d}{2}, \frac{4 (E + i \gamma)^2 q^2}{(q^2 + 4 i E \gamma)^2}
        \right) + \bigl\{ E \mapsto -E \bigr\}
      \Biggl) \\
      -\frac{2 \Gamma^2(d/2)}{\Gamma(d-1)}
      \frac{q^{2-d} \theta(q - 2E)}{\bigl[ (q^2 - 4 E^2) (q^2 + 4 \gamma^2) \bigr]^{(3-d)/2}}
    \Biggr].
\end{multline}

We have thus established exact expressions for the polarization operator for any values of the parameters $E$, $\gamma$, and $q$ and for arbitrary $d$. For the computation of diagrams in the main text, we are mostly interested in the 3D limit. More specifically, deviations from $d = 3$ should be taken into account only in the limit $q \gg E$ where dimensional regularization is applied. This means that we can substitute $d = 3$ in Eqs.\ (\ref{IRRAA}) and (\ref{IRA}) everywhere except the last terms where we keep $q^{d-3}$ asymptotic behavior. This yields the following simplified versions of the polarization operators:
\begin{subequations}
\label{Iarctan}
\begin{align}
 I_{RA}(q)
  &= \frac{\pi}{2q}\left( \frac{\pi}{2}\: q^{d-3} - E^{d-3} \arctan \frac{2\gamma}{q} \right), \label{IRAarctan} \\
 I_{RR/AA}(q)
  &= \frac{\pi}{2q}\left( \frac{\pi}{2}\: q^{d-3} - E^{d-3} \arctan \frac{2\gamma \mp 2iE}{q} \right). \label{IRRarctan}
\end{align}
\end{subequations}

\bibliography{articles}

\end{document}